# Simplications: Why and how we should rethink data of/by/for the people in smart homes and its privacy implications

A participatory research project description / position paper

**Albrecht Kurze* and Alexa Becker***

* Chemnitz University of Technology, Chair Media Informatics, Straße der Nationen 62, 09111 Chemnitz, Germany, albrecht.kurze@informatik.tu-chemnitz.de

** Anhalt University of Applied Sciences, Chair Human-Computer Interaction, Bernburger Straße 55, 06366 Köthen (Anhalt), Germany, alexa.becker@hs-anhalt.de

## Motivation

More and more smart devices enter our homes [Te20]. Often these devices come with a variety of sensors, mostly simple sensors, e.g., for light, temperature, humidity or motion. And they all collect data. While it is data of the home environment it is also data of domestic life in the home [Fi16, To16]. Thus it is data of the people and by the people in the home capturing their presence, arrival and departure, typical domestic activities, bad habits, health status etc. Based on previous as well as ongoing research we know that people are actually able to make sense of simple sensor data [Fi16, To16, Ku20] and that they will make use of it for their own purposes [Ku20].

Simple sensors, when critically reflected, are often only "simple" in a technical sense. The unreflected design and use of these sensors can easily lead to unintended implications, i.e. for privacy [Mo17]. However, it may not even need a Big Brother or data experts or AI to make the data of these sensors sensitive, e.g., if used for lateral surveillance within families [Ri18]. Often unintended but wicked [KB21] implications emerge despite good intentions, such as improving efficiency or energy saving through collecting sensor data [Ku20, De19, Be23].

Thus sensor data from the home is actually data of/by/for the people in the home.

First, we explain how this might have relevance across scales of community of people - not only for the domain of the home but also in broader meaning. Second, we relate our previous as well as ongoing research in the domain of smart homes to this topic.

## Data of/by/for the people across scales

We are interested to learn across the scales of communities of people - learning from the small scale community (in the home / individuals and families) for the bigger scale (city / society) but also vice versa. In terms of data governance we see a lot of potential common topics on different scales but with similar challenges, e.g.:

**Community**: Effectively a home / a household is a small community of household / family members. Or at least it should be, to collectively, like a bigger community, decide on how and under which conditions and for what purposes data should be collected, used and reused.

**Shared space:** A home is a shared space, shared among different dwellers, with some rooms shared and some very private. Since there are good reasons to collect, analyze and use data [Be18, De19, Ku20] it is also necessary to acknowledge boundaries between different types of space and data usage – such as walls, doors, fences and curtains do in the real world.

**Power imbalances:** Exclusive right to install sensors, access to data, higher data literacy and the ability to use the data for own purposes can consolidate or reinforce existing power imbalances in household communities. Often it is the man in a partnership that is a bit more tech savvy, interested and thus the person that installs and manages the smart device - however it might also be the wife. The point here is: It is typically a single person that easily becomes a data dictator in the home.

**Mighty 3rd party stakeholders:** While participation is not only relevant in the home - it is of extraordinary importance there - caused by complexity of context, i.e. when considering different roles and stakeholders. This might include the creators of smart technology, often tech giants, that might hold some exclusive rights of handling in their clouds the data of and by the people from home. It might also include the landlord, since a rented flat is still his property and he might have a plausible interest to have data of the building / flat but without having data of the dwellers' life. And in the end it might be "the government" or its authorities (e.g., intelligence services and police), using the data from the home, by and of the people, for their purposes, e.g. to check compliant behavior, law-abidance or to solve crimes.

**Implications and approaches independent of scale:** Since the challenges across the scales have common topics it is viable to expect the same for implications (e.g., for privacy) and for approaches to address them. This might also include data governance principles. Limited power, multi-eye principle, transparency and a system of checks and balances might work across the scales. Trusted third parties in the role of data custodians are such a promising approach, e.g. for access of third parties to data, allowing only predefined analytics and very specific use of data.

# Project Simplications

**Background:** "Simplications" is a research project (2023 - 2026) funded by the German Federal Ministry of Education and Research, FKZ 16KIS1868K. It consists of an interdisciplinary network of Chemnitz University of Technology, Anhalt University of Applied Sciences and the Saxony Consumer Advice Centre.

## Our participatory approach

"Simplications" is about simple, connected IoT sensors in the smart home and their implications. Together with consumers, we will conduct participatory research for privacy by co-design on possible implications of using seemingly simple but networked sensors in the home. We not only want to understand the WHY of implications, but also HOW we can address them in design and in actual use.

**Sensorkit for data collection and data access**: The Sensorkit used for our field studies is a self-contained system with simple sensors for temperature, movement, light and humidity based on [Be18, Ku22]. In Simplications, we expanded the system to include an air quality sensor (concentration of volatile organic compounds and $CO_2$) and a loudness sensor. The sensors are connected to a Raspberry Pi via Bluetooth Low Energy or WLAN, which records the measurement data from the sensors and stores it in a local database. The participants in the study can access the recorded data within their household using a tablet and thus independently explore visualizations of the recorded sensor data.

**Cultural Probe field studies:** The Sensorkits will be installed in households for a period of ten to 14 days. On the first day, the Sensorkit is handed over to the participants and is set up in the home with the researchers. In addition, a set-up interview is conducted and field notes are taken. A task booklet is provided for each day, which invites participants to explore the data and later also to change the position of the sensors. On the last day, the Sensorkit is dismantled again and a dismantling interview is conducted.

**"Guess the Data" data-driven group discussions:** Around ten days after the dismantling, the participants from two or three households come together for the group format "Guess the Data" [Ku20]. We then present anonymised excerpts of the collected sensor data to our participants. They then jointly interpret what the data shows, which household it comes from and what possible implications it has.

## Privacy by Co-Design

Our findings will be used to develop and evaluate media and interventions for digital education that will contribute to the informed design and use of sensor data applications in the home.

Since we are dependent on knowledge generated from the studies together with data by people, but we will also reflect this knowledge back into the living environment of the people, we call this **Privacy by Co-Design** and thus add the element of participation to the approaches Privacy by Design and Privacy by Default.

**Implications for Use**
What can be learnt from the emergence of privacy implications for the use of smart technology? From this, information and educational materials will be developed as handouts for the responsible selection and use of smart technology as well as for data release and sharing, including practical tips and tricks.

**Implications for Design**
What can be learnt from the emergence of privacy implications for the development of smart technology? From this, tips for technology development are derived in order to make data collection and interpretation privacy-friendly right from the beginning. This might also include data governance principles.

## About us

**Albrecht Kurze:** Albrecht Kurze is a post-doctoral researcher at the chair Media Informatics at Chemnitz University of Technology and one of the principal investigators in Simplications. With a background in computer science his research interests are on the intersection of Ubiquitous HCI and human centered IoT: How do sensors, data and connectedness in smart products and environments allow for new interactions and innovation and how do we cope with the implications that they create, i.e. for privacy in the home.

**Alexa Becker:** Alexa Becker is a research associate at Anhalt University of Applied Sciences and the University of Leipzig with an interdisciplinary background in literature, art, media and computer science. Her research interests are participatory tools and methods, tangibles and micro-interactions especially in the context of the smart home. During the last years she conducted multiple co-design workshops for the smart home in different settings with a variety of people.